\documentclass{iau} 
\usepackage{graphicx}
\usepackage{hyperref}

\title[Galactic Archeology with 4MOST] 
{Galactic Archeology with 4MOST}

\author[Sofia Feltzing \& et al.]   
{S.~Feltzing$^1$,
T.~Bensby$^1$, M.~Bergemann$^2$, C.~Chiappini$^3$, N.~Christlieb$^4$, M.-R.~Cioni$^3$, A.~Helmi$^5$, M.~Irwin$^6$, I.~Minchev$^3$, E.~Starkenburg$^3$, R.~de Jong$^3$}

\affiliation{$^1$Lund Observatory, Department of Astronomy and Theoretical Physics, \\ Box 43, SE-221 00, Lund, Sweden, email: {\tt sofia@astro.lu.se} \\[\affilskip]
$^2$ Max-Planck-Institut f\"ur Astronomie, Heidelberg, Germany\\[\affilskip]
$^3$ Leibniz-Institute for Astrophysics Potsdam, Berlin, Germany\\[\affilskip]
$^4$ ZAH, Landessternwarte, Heidelberg, Germany\\[\affilskip]
$^5$ Kapteyn Astronomical Institute, Groningen, The Netherlands\\[\affilskip]
$^6$ Institute of Astronomy, Cambridge, UK}

\pubyear{2018}
\volume{334}  
\jname{Rediscovering our Galaxy}
\editors{C. Chiappini, I. Minchev, E. Starkenburg, \& M. Valentini, eds.}
\begin{document}

\maketitle

\begin{abstract}
4MOST is a new wide-field, high-multiplex spectroscopic survey facility for the VISTA telescope of ESO.  Starting in 2022, 4MOST will deploy more than 2400 fibres in a 4.1 square degree field-of-view using a positioner based on the tilting spine principle. In this contribution we give an outline of the major science goals we wish to achieve with 4MOST in the area of Galactic Archeology. 
The 4MOST Galactic Archeology surveys have been designed to  address long-standing and far-reaching
problems in Galactic science. They are focused on four major themes: 1) Near-field cosmology tests, 2)  Chemo-dynamical characterisation of the major Milky Way stellar components, 3) The Galactic Halo and beyond, and 4) Discovery and characterisation of extremely metal-poor stars. 

In addition to a top-level description of the Galactic surveys we provide information about how the community will be able to join 4MOST via a call for Public Spectroscopic Surveys that ESO will launch. 
\end{abstract}

\firstsection 

\section{A brief introduction to the 4MOST facility}

The 4MOST instrument will be mounted on the 4-meter VISTA telescope that is operated by European Southern Observatory (ESO). Together, the instrument and the telescope make up the 4MOST survey facility. The 4MOST facility will provide a fully integrated system including operations and survey strategy. Finally, importantly 4MOST is a `one design fits many science cases'. With this in mind 4MOST has been designed to efficiently run all science cases, Galactic and Extragalactic, in parallel at all times. 

The 4MOST consortium has undertaken to operate a 5 + 5 year public spectroscopic survey, with the option of a third five year survey after the completion of the first two. During the first survey period the consortium surveys will take 70\,\% of the available fibre hours whilst the remaining 30\,\% will be open to the community (see also Sect.\,\ref{sect:community}). 

\begin{figure}[b]
\begin{center}
\includegraphics[width=11cm]{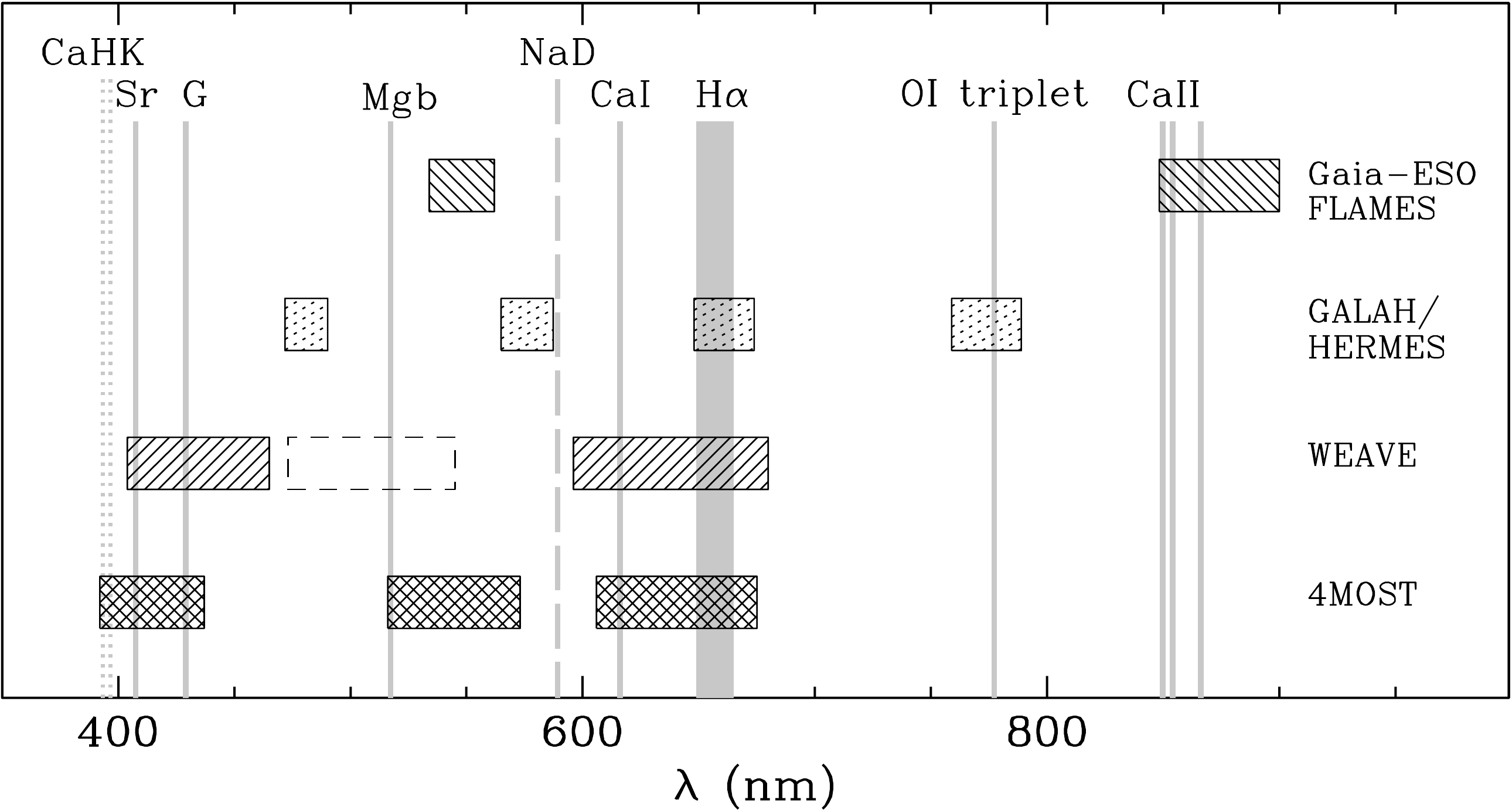} 
 \caption{The wavelength coverages of four different spectrographs. At the top a number of spectral features of importance are indicated. The first set of wavelength coverages shows the two FLAMES settings chosen by the \textit{Gaia}-ESO Survey, \cite{gilmore}, for their Milky Way field sample in the disk and halo. These were chosen based on a combination of spectral information available in the wavelength range and the throughput of the setting. Below that are the four passbands of the HERMES spectrograph that is used by GALAH, \cite{martell}. WEAVE has three passbands, however, at any given time only two are used. The choice is to have either the blue or the green band, \cite{dalton}. At the bottom we show the three passbands of the 4MOST HRS. They will always be used simultaneously.}
   \label{lambda.fig}
\end{center}
\end{figure}

\noindent
4MOST consists of three spectrographs, two low resolution spectrograph (LRS)   and one high resolution spectrograph (HRS). Each of the spectrographs are fed by 800 fibres, meaning that in total 2400 objects can be observed simultanously. The resolution of the LRS is around 6\,500 and for the HRS its about 20\,000. 
The wavelength coverage of the the LRS is  $370-950$\,nm.
 The wavelength coverage for the HRS it is split over three passbands: $392.6-435.5$, $516-573$, and $610-679$\,nm. These are shown in Fig.\,\ref{lambda.fig} where the 4MOST coverage is compared with those of other major survey instruments operating at high resolution in the optical wavelength region. How the exact wavelength coverage for the three passbands were decided upon was  first discussed in 
\cite{caffau} and further detailed in \cite{hansen} (blue band) and \cite{ruchti} (green and red bands). The definition of the edges of the blue band is driven by the need to measure the abundances of 20 elements in the very metal-poor stars in the halo (see Sect.\,\ref{halo.sect}). The green and red bands are designed to include not only lines for important elements but also to allow for a self-consistent determination of all stellar parameters, for example $T_{\rm eff}$ and $\log g$, from the spectra themselves. For these reasons both the Mgb triplet and the H$\alpha$ line are included as well as the Ca\,{\sc i} line at 616.2\,nm, which is a gravity indicator. \cite{ruchti} provides a detailed discussion of the different lines used for determination of stellar parameters.
 
The LRS covers the full spectral range of $370-950$\,nm. This is very similar to many of the other survey instruments including WEAVE (\cite{dalton}), DESI ({\cite{desi}), LAMOST, and SEGUE. Further details on exposure times and how the resolution varies across each passband can be found in \cite{dejong} and \cite{dejong2}. See also see Fig.\,\ref{snr.fig}.  The preliminary designs of the two spectrographs  are further described in \cite{laurent} and \cite{seifert}. A complete list of papers from the 4MOST consortium can be found at \url{https://www.4most.eu/cms/publications/}.

From the point of the view of Galactic Archeology the resolution and signal-to-noise ratio (SNR) for the LRS are designed to meet our need to match the uncertainty in \textit{Gaia} proper motions at the faintest magnitudes. 
This leads to the requirement of a radial velocity uncertainty of less than 2\,km\,s$^{-1}$. For the LRS 
we reach a SNR of 30 per {\AA} in 2 hours for mag$_{\rm AB} = 18.5$ with bright sky (for elemental abundances) and a SNR of about $7-10$ per {\AA} in 2 hours for mag$_{\rm AB} = 21$, which is fainter than the \textit{Gaia} limit and ensures that we can measure accurate enough radial velocities any star in the \textit{Gaia} catalogue (see Figs.\,\ref{rv.fig} and \ref{snr.fig}).   

Current estimates show that we can obtain a spectrum with SNR of 140 per {\AA} in 2 hours with the HRS for  mag$_{\rm AB} = 15.7$, which meets the requirement from the surveys that need a high precision in their determination of elemental abundances (see Fig.\,\ref{snr.fig}).

\section{Galactic Archeology with 4MOST}

\begin{figure}[t]
\begin{center}
\includegraphics[width=13cm]{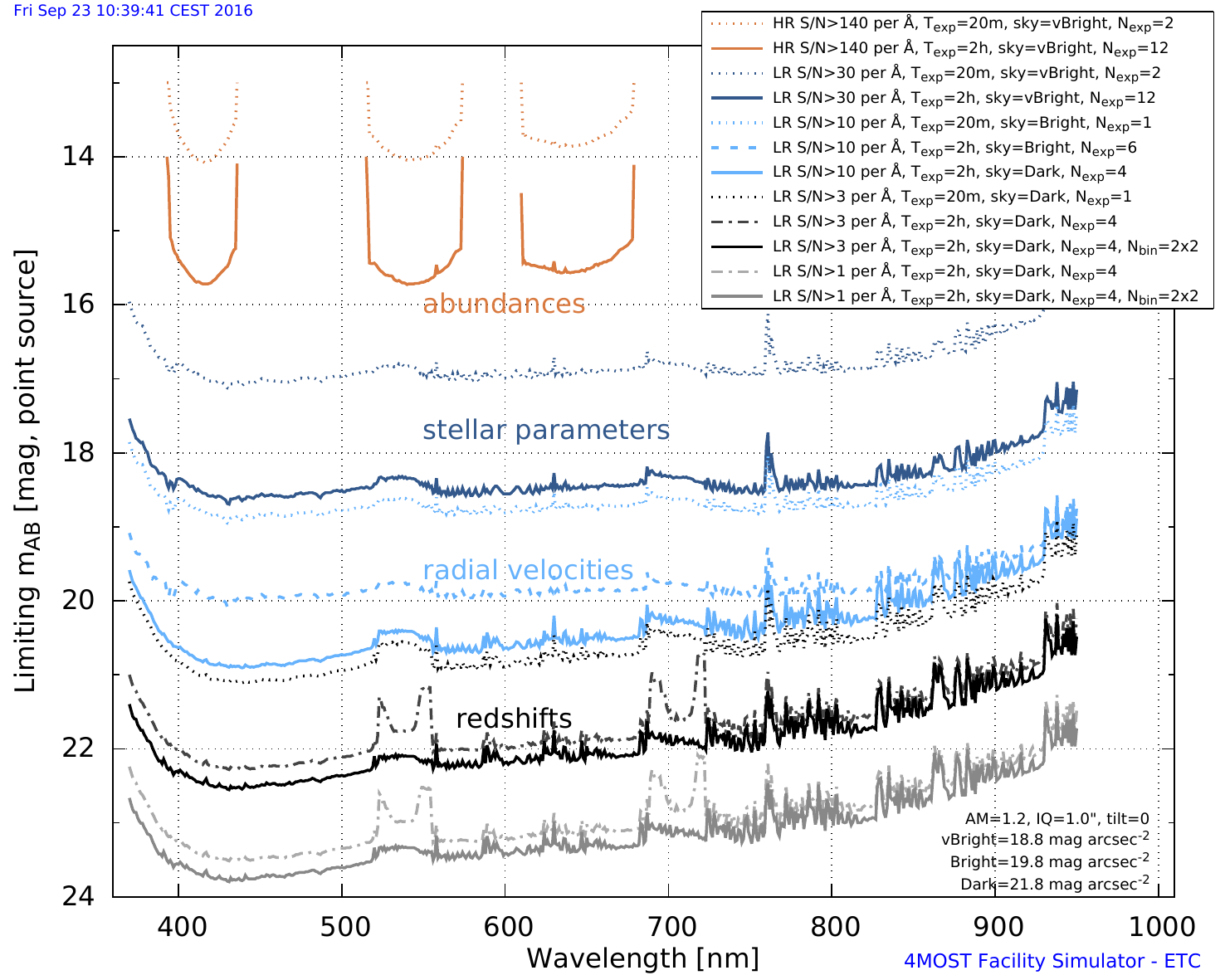} 
 \caption{Signal-to-noise ratios as a function of object magnitude (in mag$_{\rm AB}$) and exposure times for 4MOST as derived from the 4MOST Faclity Simulator (4FS). }
   \label{snr.fig}
\end{center}
\end{figure}

\begin{figure}[t]
\begin{center}
\includegraphics[width=10cm]{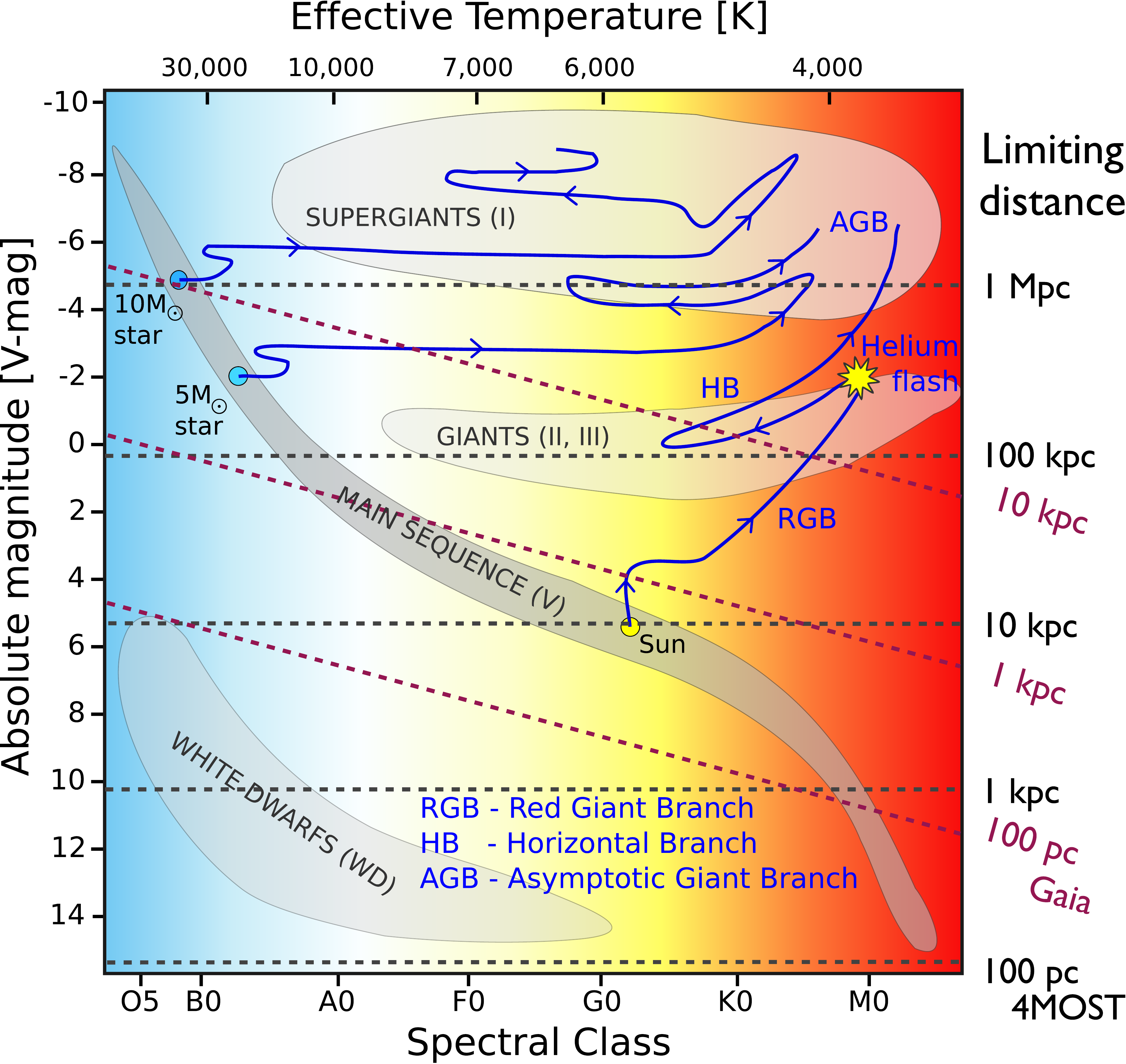} 
 \caption{This Hertzsprung-Russell diagram  illustrates to which distance for a given type of star 4MOST can obtain a radial velocity for  (dashed black lines). It also shows the  limiting distances for radial velocity measurements with \textit{Gaia} (dashed maroon diagonal lines).
}
   \label{rv.fig}
\end{center}
\end{figure}

The 4MOST Galactic Archeology surveys have been designed to  address long-standing and far-reaching
problems in Galactic science, see for example \cite{turon} for an in depth analysis. We have chosen to focus on four major themes: 
\smallskip

\begin{itemize}

\item Near-field cosmology tests 

\item Chemo-dynamical characterisation of the major Milky Way stellar components 

\item The Galactic Halo and beyond

\item Discovery and characterisation of extreme metal-poor stars 
\end{itemize}
\medskip

Although the Galactic Archeology surveys in 4MOST are designed around specific science goals we also want to stress the exploratory nature of  large surveys such as those described in this contribution. 

Our surveys will be the first large surveys that can take the full \textit{Gaia} data into account when selecting targets. This opens for new approaches in survey design. One example is our ability to avoid large foreground contamination by M dwarfs in samples studying red giant stars with spectral classes similar to the M dwarfs, and to also choose objects with small parallax errors

With its wide field, large number of fibres, and continuous observing mode (\cite{dejong}) 4MOST is built to  obtain large numbers of spectra for key components of the Milky Way in an efficient way. Over the initial five year survey we will obtain up to 20 million spectra of stars in the disk, bulge and halo with the LRS and up to 4 million spectra of disk, bulge and halo stars with the HRS.

\subsection{The Milky Way stellar disk and bulge}

As the Milky Way is the only spiral galaxy for which large numbers of individual stars can be resolved and studied in great detail, and therefore it may serve as a benchmark galaxy when constraining theoretical models of galaxy formation and evolution, it is  important to have as a complete picture as possible of the Milky Way ({\cite{freeman}, \cite{bhg}).

The 4MOST consortium plans two complementary surveys of the Milky Way stellar disk and bulge, one with the LRS and one with the HRS. The main focus of the LRS survey is the chemo-dynamics whilst the HRS survey aims at studying finer details in abundance space and couple that to the large scale picture painted by the LRS survey.

\subsubsection{Surveying the disk and bulge with the LRS}

The mechanisms of the formation and evolution of the Milky Way are encoded in the location, kinematics, and chemistry of its stars. With the 4MOST LRS survey of the disk and bulge we aim at studying kinematical and chemical substructures in the disks and bulge out to larger distances and greater precision than conceivable with \textit{Gaia} alone or any other ongoing or planned survey. 

The LRS survey of the disk and bulge is designed to 
(1) better understand the current Milky Way disk structure and dynamics (bar, spiral arms, stellar radial migration); (2) study the chemo-dynamics of the disk, which will allow us to recover the disk evolutionary history; (3) better understand the formation of the Milky Way bulge/bar using both chemical and dynamical information; (4) study the inner-disk/bulge and disk/halo interfaces by covering a large area and ensuring high quality chemical and dynamical information. The large number of targets and area covered by this survey will offer legacy samples, with the discovery of rare objects (e.g. metal-poor bulge stars), suitable for higher-resolution follow-ups

A requirement from this survey is to obtain radial velocities with accuracies matching those of the \textit{Gaia} parallaxes which leads to a $\sigma_{\rm RV} =2$\,km\,s$^{-1}$ as this will enable full exploitation of the \textit{Gaia} data for the purpose of studying the dynamics of the Milky Way. Another requirement is to obtain stellar metallicities and elemental abundances with a precision of 0.2\,dex. With these, it will be possible to distinguish and characterise different stellar populations.

To ensure that all science questions can be answered the survey is divided into four parts: the extended solar neighbourhood (with the very best parallaxes), the dynamical disk, the chemo-dynamical disk, and the inner-disk-bulge transition. The stars are selected in the simplest possible way using parallaxes, parallax errors, and magnitudes from \textit{Gaia}, thus avoiding metallicity or age biases. When necessary, for example to not contaminate the sample by foreground stars, a colour cut is also imposed. 

Of crucial importance for the LR chemo-dynamical survey (as well as for the HR Galactic surveys) and, in the construction of a training sample for 4MOST (see Sect.\,\ref{sect:an}) for data driven analysis of the spectra, will be an extensive follow-up of seismic targets from CoRoT, Kepler, K2 and PLATO\,2.0 (see Miglio \textit{et al.} 2017, Valentini \textit{et al.} 2017).

This survey will cover the entire stellar disk observable from the Southern hemisphere, approximately 20\,000 square degrees.
Keeping in mind that 4MOST operates in the visual, objects with too high extinction will be avoided (roughly $|b|<3^{\circ}$).  In the areas of interest the stellar density varies considerably but is always high, meaning that we can fill all fibres. An important aspect of this survey is the need to obtain contiguous sky coverage, this is particularly important for the dynamical studies.

\subsubsection{Surveying the disk and bulge with the HRS}

To put stringent constraints on the physics driving the formation of our own Galaxy, and disk galaxies in general, we need large, statistically significant, and unbiased samples of stars with high-precision abundances tracing all relevant nucleosynthetic channels, for the different constituents of the Milky Way. This survey will provide these data.

The 4MOST HRS survey of the disk and bulge is the first large high-resolution spectroscopic survey that can, and will, utilise from the start the full potential of the wealth of information that the \textit{Gaia} mission will provide, including prior distance and orbital information along with metallicity estimates in all Galactic populations. The main science goal is to provide decisive observational input for understanding the formation and evolution of the two main stellar populations of the Milky Way, the disk and the bulge. The science case is built around three central questions: (1) What is the Galactic bulge? (2) Does the Milky Way have two disk populations? (3) What is the merger history of the Milky Way? Solutions and answers to these questions will be acquired through a Galactic exploration of a unique densely sampled map of the chemo-dynamical structure of stars in different regimes of abundance space; from the bulge to the inner disk, through the solar neighbourhood, and out into the outer disk regions. The map will be based on high-resolution spectra for up to 4 million stars, for which we will deliver high-precision abundances for more than 15 chemical elements (the goal is to obtain elemental abundances with a precision of up 0.03\,dex) and complementary information such as stellar orbits and stellar ages, providing important constraints on the pathways leading to the present-day Galactic architecture. 

Stars will be selected based on a number of criteria, making the best use of the \textit{Gaia} data. Parallaxes and their errors, magnitudes, colours and in some cases additional information such as metallicity will be used to ensure probing all populations in the disk and Bulge. Stars will be selected with $ 14<V<17$, where the deeper observations will take place in special extragalactic drilling fields.
The survey is all sky, i.e. for 4MOST (being placed in the South) this is $0^{\circ} < {\rm R.A.} <360^{\circ}$ and $-70^{\circ} < \delta < +20^{\circ}$, which translates to roughly $-90^{\circ} < b < +90^{\circ}$ and $-45^{\circ} < l < 180^{\circ}$ in Galactic coordinates.

\subsection{The Galactic halo: LRS and HRS surveys}
\label{halo.sect}

Studies of the Galactic halo is a key science driver for Galactic surveys in 4MOST. The surveys are designed to tackle a number of questions including:

\begin{itemize}
\item Determination of the density profile, shape and characteristic parameters of the dark matter halo of the Milky Way 
\item Quantification of the amount of kinematic substructure as function of distance and location on the sky 
\item What is the shape of the low-metallicity tail of the halo MDF?
\item Does there exist a sharp cutoff of the halo MDF?
\item Is the MDF constant throughout the halo?
\end{itemize}

To be able to answer these and similar questions 4MOST will obtain LRS spectra for more than $1.8$ million stars in the halo 
(goal 3 million). The selection of the targets will be
based on \textit{Gaia}  magnitudes but various colour cuts and the usage of parallax and possibly also proper motions to weed out foreground contaminants is essential. We will target all halo giants with $15 < V < 20$ in a contiguous area covering more than 10\,000 square degrees. 

To obtain more detailed answers to some of the questions we require observation with the HRS, which is done in parallel with the LRS observations and hence can cover the same area but focusing on brighter targets ($12 < V < 16)$. For this survey we will make use of additional information and select stars that are already known to be metal-deficient with the aim to have observed 100\,000 genuine halo stars sparsely sampling 14\,000 square degrees
by the end of the first five year survey. To obtain such a large number of genuine and very metal-poor stars we will need to survey of the order of 1.2 million stars as not all stars that have tentatively been identified as belonging to the halo population(s) will be; meaning that many stars will belong to the thin or thick disk populations. The spectra of these stars will be shared with the disk surveys.
 
\subsection{Beyond the halo}

The 4MOST consortium is developing an ambitious survey of the Magellanic Clouds covering 1000 square degrees
with stars brighter than $r = 20$, enabling the study of up to 500\,000 targets across the two galaxies. The main aim is to obtain radial velocities to get the full 3D motions of the stars in the clouds, to obtain iron abundances to derive better ages from the colour magnitude diagrams and from the period-luminosity relations, and to obtain chemo-dynamics to study the formation and evolution of the two galaxies. An important part will be dedicated follow-up for large numbers of new structures found in the \textit{Gaia} data, for example the RR Lyrae variables that extend over much larger areas than previously known (e.g., \cite{belokurov} and \cite{deason}).

Thanks to the high stellar densities in the central 50 and 200 square degrees
in the SMC and LMC, respectively, all fibres can be dedicated to observations in the Clouds whilst in the outer regions the survey will coexist with other surveys, both Galactic and Extragalactic. This provides a nice example of the power of 4MOST to cater to surveys with different target densities.

\section{Analysis of stellar spectra}
\label{sect:an}

The 4MOST consortium is building its own pipeline(s) for the analysis of the stellar spectra. With a rate of more than 10\,000 spectra per night the pipeline needs to not only provide a robust analysis, but must also be quick and have a reliable performance. Our team has extensive experience in the analysis of stellar spectra in past and on-going spectroscopic surveys (e.g., Gaia-ESO, RAVE and GALAH) as well as in-depth knowledge about the analysis of stellar spectra including 3D stellar model atmospheres and non-LTE analysis of spectra (Bergemann \textit{et al.} 2012, Lind \textit{et al.} 2012). Currently we are benchmarking various methods (including both physical modelling of stellar spectra as well as data-driven approaches) and working on how to best implement the results from 3D modelling of stellar atmospheres and our better understanding of deviations from LTE in our analysis. 

To validate our spectral analysis we will observe a number of well studied stars such as the \textit{Gaia} benchmark stars (Heiter \textit{et al.} 2015, Pancino \textit{et al.} 2017), asteroseismic targets (see, e.g., discussions in Martell \textit{et al.} 2017, Pancino \textit{et al.} 2017), and include other suitable fields for cross-calibrating our surveys with those already done. 
 Of key importance for the large Galactic surveys is the use of  high-quality training samples consisting of significant numbers of stars that can be used in data-driven analysis (e.g., Ness \textit{et al.} 2016). Combined with the large spectral coverage of our spectrographs this can lead to high quality
individual elemental abundances even for  spectra with a resolution of $R \approx 6\,500$.

Work on these topics will intensify in the coming year.
 
\section{4MOST community surveys, operations and timeline}
\label{sect:community}

 The 4MOST consortium is currently developing the consortium surveys. The community will be invited to put forward proposals for community surveys in 2019. This means that the consortium surveys must be mature in their design when the Call for Letters of Intent will be issued by ESO. ESO's Public Spectroscopic Survey Panel (PSSP) will select the community proposals that will be invited to submit full survey proposals. These will then be merged with the 4MOST consortium surveys to form the 4MOST five-year Survey plan. The final decision on which surveys that will be selected rests with ESO's director general, who is advised by the OPC and the PSSP.

Those community surveys that are selected will join the 4MOST science team in order to define the final survey strategy. A period  of intensive work on this is scheduled for 2021. The 4MOST consortium has agreed with ESO to reduce all spectra for all surveys (data level 1 products). \cite{walcher} provides more information on the observing preparations and data-flow. Each survey will be responsible for producing level 2 data products (e.g., stellar parameters, and abundances of chemical elements) derived from the spectra provided to them.
The 4MOST consortium will offer community surveys the possibility to join our efforts with regards to data analysis pipelines, for example the pipeline for elemental abundances. If a community survey wishes to join such an effort they must provide manpower for operating the pipeline. 

\noindent
{\bf Timeline} The current plan for 4MOST has the publication of the consortium survey White Papers set for December 2018 with the Call for Letters of Intent for the community following in May 2019 with the aim to have the combined survey program approved by the ESO Director General in March 2021 after which the community surveys will join the 4MOST science team. The science team will work on the joint survey plan with a deadline of May 2022 and the start of the survey is scheduled for September 2022.

4MOST will have its final design review in early 2018 after which the exact dates in this timeline might be adjusted.

\acknowledgement
T.B. and S.F were supported by the project grant `The New Milky Way' from Knut and Alice Wallenberg Foundation. C.C. acknowledges support from German Research Foundation (DFG) grant CH1188/2-1 and from the ChETEC COST Action
(CA16117). N.C. is supported by Sonderforschungsbereich SFB 881 `The Milky Way System' (subprojects A4 and A9) of the German Research Foundation (DFG). E.S. gratefully acknowledges funding through a German Research Foundation (DFG) Emmy Noether grant.

\end{document}